\documentclass[aip,jcp,reprint,floatfix]{revtex4-1}

\usepackage[pdftex]{graphicx}
\DeclareGraphicsExtensions{.pdf,.png}

\usepackage{grffile}

\usepackage{hyperref}
\hypersetup
{
pdfmenubar=true, 
pdfhighlight=/O, 
pdfpagelayout=SinglePage, 
pdffitwindow=true, 
colorlinks=true, 
linkcolor=blue, 
citecolor=blue, 
urlcolor=blue 
}

\usepackage{amsmath, amsthm, amssymb, amsfonts, amsbsy}
\usepackage{mathtools} 
\usepackage{bm}

\renewcommand*{\v}[1]{\boldsymbol{#1}}
\newcommand\bnabla{\boldsymbol{\nabla}}
\newcommand\bcdot{\boldsymbol{\cdot}}
\newcommand{\grad}{\bnabla} 
\renewcommand{\div}{\bnabla\bcdot} 
\newcommand{\m}[1]{\v{#1}}

\renewcommand{\d}{\mathrm{d}} 

\newcommand{\norm}[1]{\lvert #1 \rvert}

\usepackage[capitalise]{cleveref}
\crefformat{equation}{Eq. (#2#1#3)} 
\Crefformat{equation}{Equation (#2#1#3)} 
\crefrangeformat{equation}{Eqs. (#3#1#4--#5\crefstripprefix{#1}{#2}#6)} 
\Crefrangeformat{equation}{Equations (#3#1#4--#5\crefstripprefix{#1}{#2}#6)} 
\crefmultiformat{equation}{Eqs. (#2#1#3)}{ and~(#2#1#3)}{, (#2#1#3)}{ and~(#2#1#3)} 
\Crefmultiformat{equation}{Equations (#2#1#3)}{ and~(#2#1#3)}{, (#2#1#3)}{ and~(#2#1#3)} 

\usepackage[dvipsnames]{xcolor}
\newcommand{\major}[1]{#1}
\newcommand{\minor}[1]{#1}

\begin{document}

\title{Exact results for sheared polar active suspensions with variable liquid crystalline order}

\author{Aurore Loisy}
\author{Anthony P. Thompson}
\author{Jens Eggers}
\author{Tanniemola B. Liverpool}
\affiliation{School of Mathematics, University of Bristol - Bristol BS8 1TW, UK}

\date{\today}

\begin{abstract}
We solve the problem of a confined sheared active polar liquid crystal film with varying amounts of polarization, but a uniformly aligned director. Restricting our analysis to one-dimensional geometries, we demonstrate that with asymmetric boundary conditions, this system is characterized, macroscopically, by a linear shear stress vs. shear strain relationship that does not pass through the origin: at zero strain rate the fluid sustains a non-zero stress. Analytic solutions for the polarization, density and velocity fields are derived for asymptotically large or small systems, and are shown by comparison with precise numerical solutions to be good approximations for finite-size systems.
\end{abstract}


\maketitle

\section{Introduction}

Swarms of bacteria \cite{Dombrowski2004,Zhang2010,Gachelin2014,Chen2017}, mixtures of cytoskeletal filaments and motor proteins \cite{Schaller2010,Sanchez2012,Wu2017}, and self-propelled colloids \cite{Bricard2013,Theurkauff2012} are all example of active suspensions \cite{Ramaswamy2010,Marchetti2013,Prost2015,Saintillan2013} consisting of anisotropic self-driven particles dispersed in a passive liquid.
Due to the orientable nature of their constituents, active suspensions can exhibit long-range orientational order and are often referred to as active liquid crystals (LCs) \cite{Ramaswamy2010,Marchetti2013,Prost2015,Julicher2018}. While active LCs can exist in ordered phases typical of liquid crystals \cite{deGennes1993book}, they fundamentally differ from their passive counterparts in that each active particle transduces free energy into systematic movement, maintaining the system out of equilibrium.

As active particles interact with each other and with their surrounding environment, they are able to collectively generate motion and mechanical stresses at scales much larger than their individual size, endowing active materials with unusual mechanical  properties.  An example is the reduction of the apparent viscosity of bacterial suspensions under shear~\cite{Sokolov2009,Rafai2010,Gachelin2013,Mussler2013,Lopez2015,Saintillan2018}.
Remarkably, upon increasing activity,  the apparent viscosity can decrease until a value of zero is achieved, giving rise to superfluid-like behaviour~\cite{Lopez2015,Guo2018}. In the last decade, rheological measurements \cite{Sokolov2009,Rafai2010,Gachelin2013,Mussler2013,Lopez2015} have shown qualitative agreement with earlier theoretical predictions \cite{Hatwalne2004,Liverpool2006,Haines2009,Saintillan2010,Saintillan2010a,Ryan2011} for the macroscopic mechanical properties of active suspensions.
Yet a more thorough understanding of the underlying mechanisms driving these systems requires a more quantitative comparison of theoretical models with experiments~\cite{Loisy2018b}. Such a comparison is becoming possible as more detailed information, such as transient rheological behaviour~\cite{Lopez2015} and velocity profiles~\cite{Guo2018}, becomes accessible experimentally.

Many active materials, whether biological or synthetic, involve head-tail asymmetric particles and can exist in a polar phase. For such system, the broken symmetry variable is the polarization vector which represents the local coarse-grained orientation of the particles. When an active polar LC is sheared, distortions in the orientation of the polarization field induces active stresses which in turn generate an extra flow (needed to maintain the stress balance). This mechanism allows the apparent viscosity (defined as the macroscopic viscosity at the scale of the system, as would be measured by, e.g., a rheometer) of active LCs to vanish or even become negative \cite{Loisy2018b}. 
While previous theoretical analyses~\cite{Giomi2010,Furthauer2012,Loisy2018b} have focused on how gradients in the \emph{orientation} can induce the active stresses that lead to unconventional mechanical behaviour, here our focus is on how variations in the \emph{magnitude} of LC order affect the mechanical properties of active LCs.

In this paper we show that a gradient in the magnitude of polarization of active LCs induces, even in the {\em absence} of variations in orientation, flows that give rise to anomalous mechanics. 
\minor{Specifically}, we examine the effect of a varying polarization magnitude on a one-dimensional confined active polar LC subjected to shear. 
We derive the analytical relation between the stress and the macroscopic strain rate, which shows in particular that this system experiences a non-zero shear stress at zero shear strain rate (and conversely, a non-zero strain rate is required to maintain a zero stress).

Despite the nonlinearities in the equations, because of a decoupling of the equations we are also able to obtain exact analytical results for the velocity and density fields as a function of the polarization field, the latter being shown to be \minor{well} approximated by \minor{asymptotic solutions in the limit of large or small systems}.
This is interesting because hydrodynamic equations for active LCs are highly nonlinear, hence analytical solutions beyond linearization approximations are rare even in the simplest geometries. As a result studies of sheared active LCs tend to be numerical \cite{Marenduzzo2007a,Cates2008,Giomi2010,Fielding2011,Tjhung2011,Furthauer2012,Loisy2018b}.

\section{Model \label{sec:model}}

We consider an active LC with the possibility of polar orientational order. At the continuum scale, its dynamics are described by a set of long-wavelength, long-time scale equations forming the now well-accepted hydrodynamic theory of active matter \cite{Marchetti2013,Julicher2018}. The relevant hydrodynamic variables are the polarization vector $\v{p}$ as well as the conserved fields, \minor{here the particle number density $\rho$ (for simplicity $\rho$ is normalized by its equilibrium value) and the momentum $\rho_m \v{u}$ where $\rho_m$ is the fluid mass density and $\v{u}$ is the fluid velocity}.

The passive contributions to the equations of motion are customarily described as arising from the nonequilibrium analog
of the free energy for a passive polar LC.
This free energy is given by \cite{Giomi2010,Marchetti2013}
\begin{equation}
	F = \int_{\boldsymbol{r}} (f_n + f_p) \; \d \v{r}
\end{equation}
\begin{equation}
	f_n = \frac{a_2}{2} \norm{\v{p}}^2 + \frac{a_4}{4} \norm{\v{p}}^4 + \frac{K}{2} \norm{\grad \v{p}}^2 + \frac{C}{2} (\rho-1)^2
\end{equation}
\begin{equation}
	f_p = B_1 (\rho-1) \div \v{p} + B_2 \norm{\v{p}}^2 \div \v{p} + B_3 \norm{\v{p}}^2 \v{p} \bcdot \grad \rho.
\end{equation}
The contribution $f_n$ is the free energy density of a nematic LC.
The first two terms control the isotropic-polar transition: they favor a
polar phase ($\norm{\v{p}}^2 = -a_2/a_4$) when $a_2<0$ and an isotropic phase ($\norm{\v{p}}^2 = 0$) when $a_2>0$. The third term describes the energy cost of deformation ($K$ is the analog of the Frank constant for passive LCs), and the last term penalizes density variations ($C$ is the compression modulus).
The contribution $f_p$ contains additional terms that break the $\v{p} \rightarrow - \v{p}$ symmetry and are allowed in a polar fluid \cite{Kung2006}.

\begin{figure}
	\centering
	\includegraphics[width=0.99\linewidth]{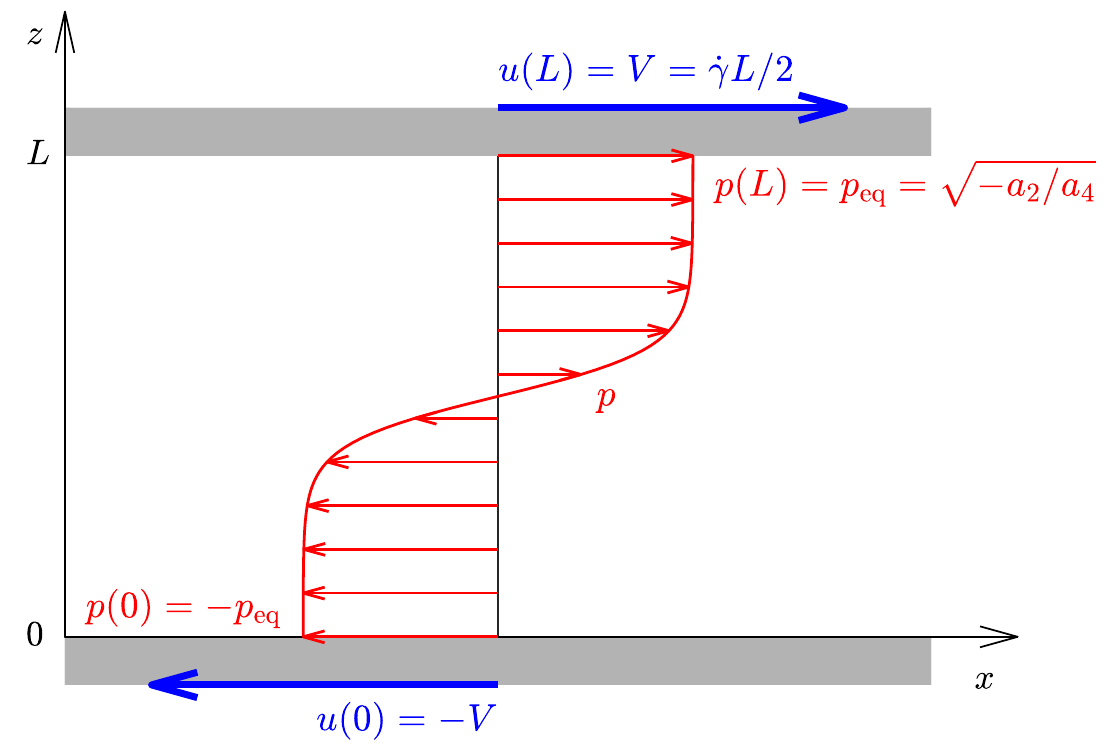}
	\caption{
	A thin film of active polar LC sheared between two moving no-slip walls. The polarization field is uniformly aligned with the walls, and only its magnitude $p$ is allowed to vary.
	\label{fig:scheme_BC_Pfield}
	}
\end{figure}

The polarization and density dynamics are governed by
\begin{equation}
	D_t \v{p} = - \beta_p \v{p} \bcdot \grad \v{p} + \lambda \m{E} \bcdot \v{p} - \m{\Omega} \bcdot \v{p} - \Gamma_{pp} \v{h} - \Gamma_{cp} \v{g}
\end{equation}
and
\begin{equation}
	D_t \rho = \div \left[ -\rho \beta_c \v{p} + \Gamma_{cp} \v{h} + \Gamma_{cc} \v{g} \right],
\end{equation}
where $D_t = \partial_t + \v{u} \bcdot \grad$, $E_{ij}=(\partial_i u_j + \partial_j u_i)/2$, $\Omega_{ij}=(\partial_i u_j - \partial_j u_i)/2$, $\v{h} = \delta F/\delta \v{p}$ and $\v{g} = \grad (\delta F/\delta \rho)$. 
The flow is assumed incompressible ($\div\v{u} = 0$) and \major{the flow field satisfies 
\begin{equation}
	\rho_m (\partial_t + u_i \partial_i) u_j =  \partial_i \sigma_{ij}.
\end{equation}}
The stress tensor is given by
\begin{equation}
	\sigma_{ij} = 2 \eta E_{ij} + \sigma_{ij}^r + \sigma_{ij}^a,
\end{equation}
where the first term is the dissipative contribution ($\eta$ is the fluid viscosity), \minor{$\sigma_{ij}^r$ is reversible contribution (as in passive LCs), and $\sigma_{ij}^a$ is the active contribution.
The reversible stress is given by}
\begin{equation}
	\sigma_{ij}^r = - \Pi \delta_{ij} + \frac{\lambda}{2} (p_i h_j + p_j h_i ) \\ + \frac{1}{2} (p_i h_j - p_j h_i),
\end{equation}
where $\Pi$ is the pressure.
\minor{The active stress is}
\begin{equation}
	\sigma_{ij}^a = \alpha \rho p_i p_j + \beta_\sigma \rho (\partial_i p_j + \partial_j p_i),
\end{equation}
where the lowest order term $\sim \alpha$ has nematic symmetry while the higher order term $\sim \beta_\sigma$ is present only in systems with polar symmetry.
In the above equations, $\beta_{c,p}$ and $\beta_\sigma \Gamma_{pp}$ have the dimension of a velocity and associated terms arise in polar systems from the self-advection of active elements along $\v{p}$.

Our geometry is similar to that used in prior work \cite{Giomi2010,Loisy2018b} and is depicted in \cref{fig:scheme_BC_Pfield}: a two-dimensional thin film of active LC of thickness $L$ is sheared between two parallel walls moving in opposite directions with velocity magnitude $V$. We allow gradients only in the direction normal to the walls. 
Due to incompressibility and wall impermeability the flow must be parallel to the wall: $\v{u} = (u(z), 0)$, and we use no-slip boundary conditions: $u(L)=-u(0)=V$. The fluid is therefore subjected to a macroscopic shear strain rate $\dot{\gamma} = \int_0^L \partial_z u \, \d z = 2 V / L$.
 
To pick out the effects of variations in the amount of LC order \minor{only},
we further assume that the polarization field is parallel to the walls and only its (signed) magnitude is allowed to vary: $\v{p} = (p(z), 0)$.
\major{Previous work has focused on the role of varying orientation with fixed magnitude, here in contrast we fix the orientation and focus on the role of varying magnitude of $\v{p}$ only. Our choice of orientation is consistent with expected boundary conditions on the walls. A nice feature of this approximation is that it makes the problem tractable analytically due to a decoupling of the equations.
Such an aligned state is physically relevant to situations where strong parallel anchoring is precribed at the walls, provided that the coupling between the polarization orientation and the local shear is negligible, that is, for systems which satisfy $\Gamma_{pp} K \gg U L$ (with $U$ a characteristic velocity scale for the flow).} 
The governing equations reduce to
\begin{equation}
	\partial_t p = - \Gamma_{pp} ( a_2 + a_4 p^2 ) p + \Gamma_{pp} K \partial_z^2 p,
	\label{eq:polarization_1D}
\end{equation}
\begin{equation}
	\partial_t \rho = \partial_z \big[ - 2 b_2 p \partial_z p + (d + b_3 p^2) \partial_z \rho \big],
	\label{eq:density_1D}
\end{equation}
\major{\begin{equation}
	\rho_m \partial_t u = \partial_z \sigma,
	\label{eq:velocity_1D}
\end{equation}
with}
\begin{equation}
	\sigma = \eta \partial_z u + (\beta_\sigma \rho + m b_2 p^2) \partial_z p + \frac{m}{2} (b_1 - b_3 p^2) p \partial_z \rho,
	\label{eq:sigma_1D}
\end{equation}
\minor{where $\sigma_{zx}$ is now simply denoted $\sigma$}, and where we have introduced $d=\Gamma_{cc} C - \Gamma_{cp} B_1$, $b_{1,2,3} = \Gamma_{cp} B_{1,2,3}$, and $m=(1-\lambda)/\Gamma_{cp}$.
At the boundaries the flux of $\rho$ across the walls must be zero, and we require that the polarization vectors at the walls are antiparallel: $p(L)=-p(0)=p_{\mathrm{eq}}=\sqrt{-a_2/a_4}$ (here we assume $a_2 < 0$).

It is interesting to note that the coupling terms in the governing equations are those which break the $\v{p} \rightarrow - \v{p}$ symmetry. For a nematic fluid ($b_{1,2,3}=0$, $\beta_\sigma=0$), the equation for $\rho$ reduces to a diffusion equation, and the expression of the stress only contains the viscous term. Therefore, in the simple configuration we consider here, a nematic active fluid would behave as an isotropic passive one.

\section{Results \label{sec:results}}

Let us consider a continuous steady solution $p(z)$ for the polarization field. Then the steady solution to \cref{eq:density_1D} is
\begin{equation}
	\rho = \frac{b_2}{b_3} \ln \left( d + b_3 p^2 \right) + \rho_0,
\end{equation}
where $d + b_3 p^2 > 0$ is assumed \footnote{otherwise this would be equivalent to a negative diffusion coefficient in the density equation, meaning that fluctuations around the equilibrium density would not remain small and that stabilizing higher order terms would need to be included in our model} and where $\rho_0$ is a constant determined by the condition $L^{-1} \int_0^L \rho \; \d z= 1$.
\minor{At steady state the shear stress is uniform across the gap ($\partial_z \sigma = 0$)}.
One can then integrate \cref{eq:sigma_1D} for a constant (unknown) $\sigma$ and obtain the velocity field
\begin{equation}
	u = \frac{\sigma}{2 \eta} (2z-L) - \xi p + \sqrt{\frac{d}{b_3}} \xi \arctan (\sqrt{\frac{b_3}{d}} p) - \frac{\beta_\sigma}{\eta} p \rho
	\label{eq:u_solution}
\end{equation}
with
\begin{equation}
	\xi = \frac{b_2}{\eta b_3} (m b_1 + m d - 2 \beta_\sigma).
\end{equation}

The (macroscopic) \minor{steady-state} flow curve $\sigma=f(\dot{\gamma})$, as would be measured by a rheometer, is then obtained from \cref{eq:u_solution} by satisfying the no-slip boundary conditions at the moving walls. One finds
\begin{equation}
	\sigma = \eta \dot{\gamma} + \sigma_{0}
\end{equation}
with
\begin{multline}
	\sigma_{0} = \frac{2 \eta}{L} \Bigg\{ \xi p_{\mathrm{eq}} - \sqrt{\frac{d}{b_3}} \xi \arctan (\sqrt{\frac{b_3}{d}} p_{\mathrm{eq}}) \\+ \frac{\beta_\sigma}{\eta} p_{\mathrm{eq}} \left[ \frac{b_2}{b_3} \ln \left( d + b_3 p_{\mathrm{eq}}^2 \right) + \rho_0 \right] \Bigg\}.
	\label{eq:sigma0}
\end{multline}
The slope $\d \sigma / \d \dot{\gamma}$ is simply the fluid viscosity $\eta$, as for an isotropic passive fluid, however the stress at zero strain rate, $\sigma_0$, is not zero\major{: the active LC has a yield stress and effectively behaves, from a rheological point of view, in a similar way to a Bingham fluid \cite{Bingham1922}}. Indeed gradients in $p$ induce reversible and active contributions to the stress which exist independently of the applied strain rate. The converse is also true: a non-zero macroscopic strain rate is required in order to maintain a zero stress. Moreover $\sigma_0$ can be of either sign, meaning that the apparent viscosity, defined as $\sigma/\dot{\gamma}$, can be negative (while passive contributions can only add to $\sigma$, active ones $\sim \beta_\sigma$ can either add or subtract to $\sigma$).

\begin{figure}
	\centering
	\includegraphics[width=0.94\linewidth]{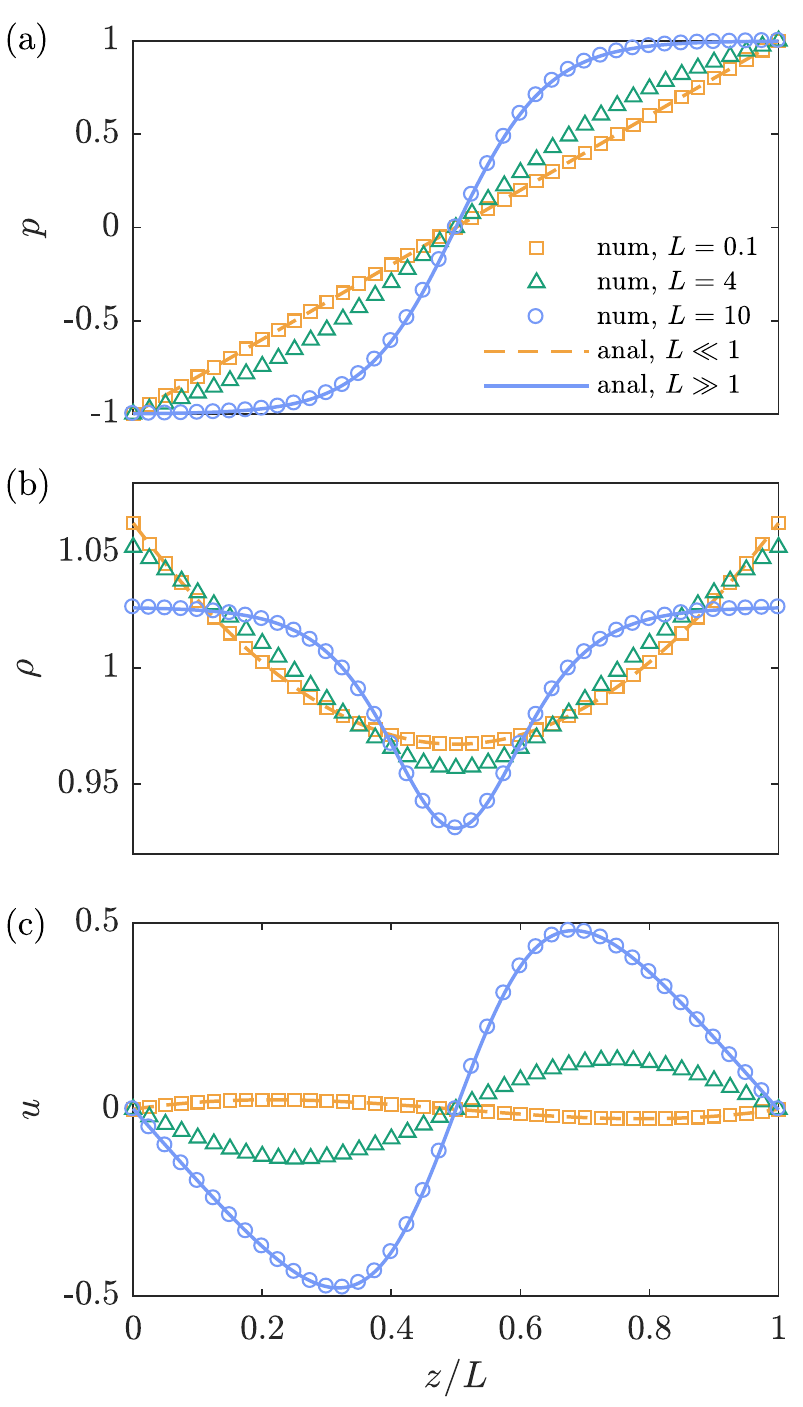} \\
	\includegraphics[width=0.94\linewidth]{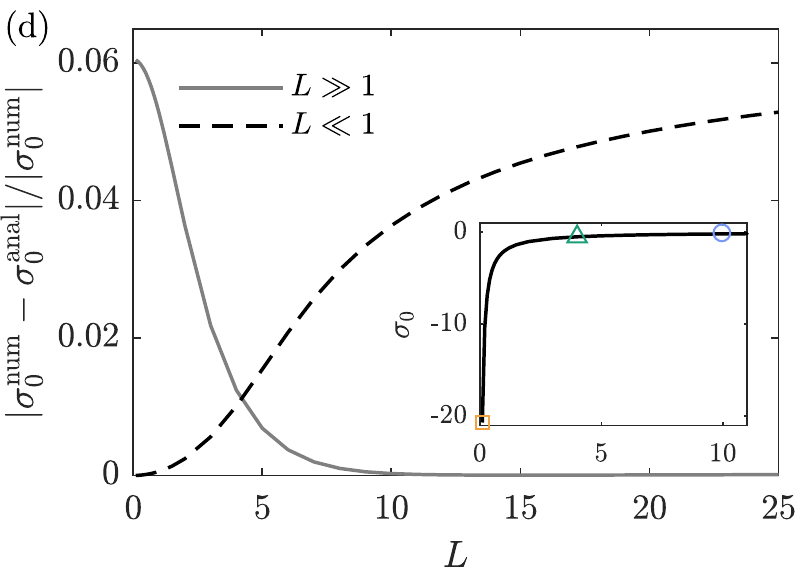}
	\caption{
	The effect of system size: (a) polarization magnitude field, (b) density field, (c) velocity field, (d) deviation from asymptotic results in the stress at zero strain rate (inset: stress). Analytical profiles and stress were obtained using \cref{eq:solp_tanh} for $L \gg 1$ and \cref{eq:solp_linear} for $L \ll 1$ (here we set for simplicity $-a_2/K=1$). Parameters are: $a_2=-1$, $a_4=1$, $K=1$, $\eta=1$, $b_{1,2,3}=0.1$, $m = -1$, $d=1$, $\beta_\sigma=-1$, $\dot{\gamma}=0$ (arbitrary units).
	\label{fig:size_effect_fields}
	}
\end{figure}

There exists, as far as we know, two limiting cases where an explicit steady analytical solution to \cref{eq:polarization_1D} can be written.
In the limit $-a_2 L^2 / K \gg 1$, a good approximation for $p$ is the solution for an infinite system \cite{Mazenko2003book}
\begin{equation}
	p = p_{\mathrm{eq}} \tanh \left[\sqrt{\frac{-a_2}{2 K}} (z-z_i) \right]
	\label{eq:solp_tanh}
\end{equation}
where $z_i$ is the (undetermined) location of the interface (which thickness decreases with increasing $-a_2/K$) between two polar phases pointing in opposite directions.
This profile results in a depletion (or accumulation, depending on the sign of $b_2$) of $\rho$ localized at the interface and in a non-uniform, non-monotonic velocity profile.

In the opposite limit $-a_2 L^2 / K \ll 1$, the solution for the polarization magnitude can be approximated by a linear profile
\begin{equation}
	p = - p_{\mathrm{eq}} + \frac{2 p_{\mathrm{eq}}}{L} z.
	\label{eq:solp_linear}
\end{equation}
Note however that this latter limit \minor{may} require that the width $L$ be comparable to the active particle size for which the validity of the hydrodynamic equations are in question and is \minor{mostly} of interest here as a bound.

In addition to these limiting cases, \cref{eq:polarization_1D,eq:density_1D,eq:sigma_1D} were also solved numerically. Our algorithm is based on second-order implicit finite difference schemes (Crank-Nicolson scheme for time integration and centered schemes for spatial discretization) with adaptive time-stepping. Time-dependent equations were solved to steady-state, starting from a linear profile for $p$ and a uniform $\rho$.
A comparison between the asymptotic solutions and the numerical ones is shown in \cref{fig:size_effect_fields}, together with additional results for an intermediate value of $-a_2 L^2 / K$. The estimate of $\sigma_0$ obtained from asymptotic expressions is remarkably accurate (deviation not greater than 1 \%) even for $-a_2 L^2 / K = O(1)$.

\section{Discussion and conclusions \label{sec:conclusions}}

We considered the minimal problem of a one-dimensional sheared active LC under confinement, with a uniform orientation of the polarization field, focusing on the effect of varying its signed magnitude $p$. Our analysis thereby complements prior studies of the same system which allowed gradients in the orientation field while keeping the magnitude of liquid crystalline order constant \cite{Voituriez2005,Giomi2010,Loisy2018b}.

As the dynamics of $p$ is not coupled to that of the density or the velocity, the uniform equilibrium solution is always stable and gradients in $p$ must be generated through boundary conditions. Here we imposed $p$ to be of equal magnitude and of opposite signs at the walls. Such asymmetric polarization at the boundaries could be realized experimentally through manipulation of the surface chemistry or architecture~\cite{Koumakis2014,Sipos2015,Yuan2017,Munoz-Bonilla2018,Hasan2013}.

The case of variable orientation leads to a rich phenomenology, including a spontaneous transition to a flowing state in the absence of external driving \cite{Voituriez2005}, and the existence of non-monotonic stress vs. strain rate flow curves \cite{Giomi2010,Loisy2018b}. 
\minor{In contrast, the case of variable polarization studied here does not yield such unusual mechanical properties:} the relationship between the stress and the macroscopic strain rate is linear, and
\minor{, for a nematic active LC, would not differ from that for an isotropic fluid.}
\minor{For a polar active LC though}, there exist elastic and active contributions to the total stress in addition to the viscous one, and the flow curve does not pass through the origin. This indicates that macroscopic stresses are present in \minor{the uniformly aligned polar active LC even in the absence of external driving.}

One of the advantages of the simple configuration considered here lies in the fact that analytical solutions can be explicitly obtained. We hope that these solutions will provide insight into the role played by gradients of liquid crystalline order, and could be used as a starting point and benchmark reference for numerical work on sheared active polar LCs, where both the magnitude and direction of the LC order parameter vary \cite{Marenduzzo2007a,Cates2008,Fielding2011}.

\begin{acknowledgments}
Part of this work was funded by a Leverhulme Trust Research Project Grant RPG-2016-147. APT acknowledges a University of Bristol undergraduate summer bursary.
TBL acknowledges support of BrisSynBio, a BBSRC/EPSRC Advanced Synthetic Biology Research Centre (grant number BB/L01386X/1).
\end{acknowledgments}

\bibliography{biblio}

\end{document}